\documentclass[pra,aps,twocolumn,showpacs,groupedaddress,superscriptaddress]{revtex4}

\usepackage{graphicx}

\usepackage{amssymb}
\usepackage{natbib}

\begin{document}




\title{Nonlinear magneto-optical effects in Ba vapor}


\author{I.\ Novikova}
\affiliation{Department of Physics and Institute for Quantum
Studies, Texas A\&M University, College Station, Texas 77843-4242}
\affiliation{Harvard-Smithsonian Center for Astrophysics,
Cambridge, Massachusetts, 02138}

\author{A.\ Khanbekyan}
\author {D.\ Sarkisyan}
\affiliation{Institute for Physical Research, Armenian Academy of
Science,
    Ashtarak-2 378410 Armenia}
\author{G.\ R.\ Welch}
\affiliation{Department of Physics and Institute for Quantum
Studies, Texas A\&M University, College Station, Texas 77843-4242}

\begin{abstract}

We report the first measurements of linear and nonlinear
magneto-optical polarization rotation on an intercombination
transition of Ba vapor ($\lambda= 791.1$~nm).
%
%
We observe a maximum polarization rotation angle in Faraday
configuration of $15$~mrad, accompanied by suppression of
absorption.  A theoretical treatment of the nonlinear Faraday
effect in the limit of a strong interacting light field is
developed.

\end{abstract}
%

\date{\today}
\maketitle

    In the last few decades, nonlinear magneto-optical effects
(NMOE) have been studied thoroughly in various experimental
configurations, and they have been used or proposed to be
used for many practical applications, including precision
metrology and fundamental symmetry tests (for a review see
Ref.~\cite{budker'02rmp}).  The majority of experiments,
however, are focused on the interaction of resonant light
with alkali atoms like Rb, Cs and Na. There are only a few
publications reporting studies of NMOE for alkaline-earth
atoms:  the study of nonlinear Faraday and Voigt effects in
Sm~\cite{drake'88,barkov'89} and the study of the nonlinear
Faraday effect in the regime of strong magnetic fields in
Ca~\cite{agarwal'97}.

    We present here the first (to our knowledge)
measurements  of nonlinear magneto-optical polarization rotation
in Ba vapor for laser radiation resonant with the $6s^2~{}^1S_0
\rightarrow 6s6p~{}^3P_1$ intercombination transition ($\lambda =
791.1$~nm).  This transition is of particular interest since the
angular momentum of the ground state is zero, and the nonlinear
effects are due to the interference of the magnetic sublevels of
the excited sublevels in a $V$ level scheme. It should be noted
that although there exists a number of theoretical calculations of
nonlinear polarization rotation in $J=0 \rightarrow J'=1$
transitions~\cite{schuller'87, schuller'92, schuller'99},
experimental results are very limited~\cite{barkov'89}.  Coherence
effects in a $V$-type level configuration have also been studied
in regards to lasing without inversion~\cite{zibrov'95,
grynberg'96, zhou'97}, but with very little experiments focused on
the dispersive properties of this system.

    The interaction of linearly polarized
light with the intercombination transition of Ba forms an ideal
$V$ scheme which cannot be realized in, for example, alkali atoms
because of their richer Zeeman structure. It has been demonstrated
that ground state coherence may be formed due to radiative
coherence transfer if the ground state is degenerate even in the
case of higher degeneracy of the excited state~\cite{akulshin'98,
taichenachev'00pra, alzetta'01}, which may obscure the observation
of the coherence between excited states.  It is also important
that the radiative width of the excited state ${}^3P_1$ is small
($\gamma_r = 2\pi \times 100$~kHz) which is true for the majority
of alkaline-earth elements.  Thus, the coherence created between
the excited state sublevels may exist long enough to produce a
noticeable nonlinear effect.

    In addition, since the frequencies of the
intercombination transition in Ba and the $D_1$ line of Rb
are very close ($\Delta \lambda \approx 3~\mathrm{nm}$) it is
logical to expect resonant enhancement of the spin-exchange
cross-section between atoms of the two species. Therefore,
the study of linear and nonlinear effects in Ba should provide
accurate information about the interaction with Rb atoms and
vice versa.  In this case Ba atoms may be used to monitor the
properties of very dense Rb vapors.

    In this paper we first present the experimental
data for saturation absorption spectroscopy of Ba atoms.
These data provide the necessary information about level
structure of various Ba isotopes, and allow better calibration
of the magnetic field.  Then we study the linear and nonlinear
Faraday effects for different experimental parameters.  In the
last section we analyze the observed experimental data using
density matrix formalism.

\section{Experimental setup}
\label{setup}
\subsection{Sapphire Ba cell}

    In this experiment we use a sealed cylindrical
sapphire cell (SC) containing pure Ba with birefringence-free
windows made of garnet (YAG) crystal. This cell allows operation
at high-temperature (up to $600^\circ$C on the windows). The
length of the cell is $L=102$~mm, with an inner diameter of
$10.7$~mm.  The residual vacuum in the SC is $\sim 10^{-3}$~Torr.

    Heating elements are placed closer to the windows
to avoid condensation of Ba vapor on the windows of the SC and to
prolong its useable lifetime.  The temperature is then measured by
two nonmagnetic thermocouples:  one is placed next to the window
so it detects the highest temperature of the SC, and the other is
located in the central part of the cell to detect the lowest
temperature.  Both the SC and the heaters are surrounded by a
thermo-isolating material (``shamot'').

    We conduct the experiment at two different temperatures
of the SC: at $490^\circ$C (temperature on the window is
$515^\circ$C), and at $540^\circ$C (temperature on the window is
$560^\circ$C).  There is no well-established dependence of Ba
vapor pressure on its temperature, as it is noted
in~\cite{nesmeyanov'63}.  Below, we give the standard function
describing this dependence, and the numerical values of the
coefficients from different studies:
\begin{equation} \label{density}
\log(p)=A+BT^{-1}+C\log T
\end{equation}
where $p$ is measured in Pa, and $T$ is in K.

\begin{tabular}{|c|c|c|c|}
  Reference & $A$ & $B$ & $C$ \\ \hline
  Hinnov \textit{et al.} Ref.~\cite{hinnov'69} &
    $8.9$ & $-8800$ & $0$ \\ \hline
  Alcock \textit{et al.} Ref.~\cite{alcock'84} &
    $17.411$ & $-9690$ & $-2.2890$ \\ \hline
  Jacob \textit{et al.} Ref.~\cite{jacob'88} &
    $9.733$ & $-9304$ & 0 \\ \hline
\end{tabular}

    We need to point out that the values of atomic
densities calculated using the various tabulated coefficients
are not consistent with each other for our experimental
temperatures.  In the following we will use the most recent
study of Jacob \textit{et al.}~\cite{jacob'88} which gives the
values of Ba atomic densities for our experiment to be $N=3.3
\times 10^{11} \mathrm{cm}^{-3}$ and $N=1.8 \times 10^{12}
\mathrm{cm}^{-3}$.  The data from~\cite{hinnov'69} give values
about $40\%$ lower, whereas those of~\cite{alcock'84} give a
Ba pressure an order of magnitude larger.
\begin{figure}
\centering
\includegraphics[width=0.70\columnwidth]{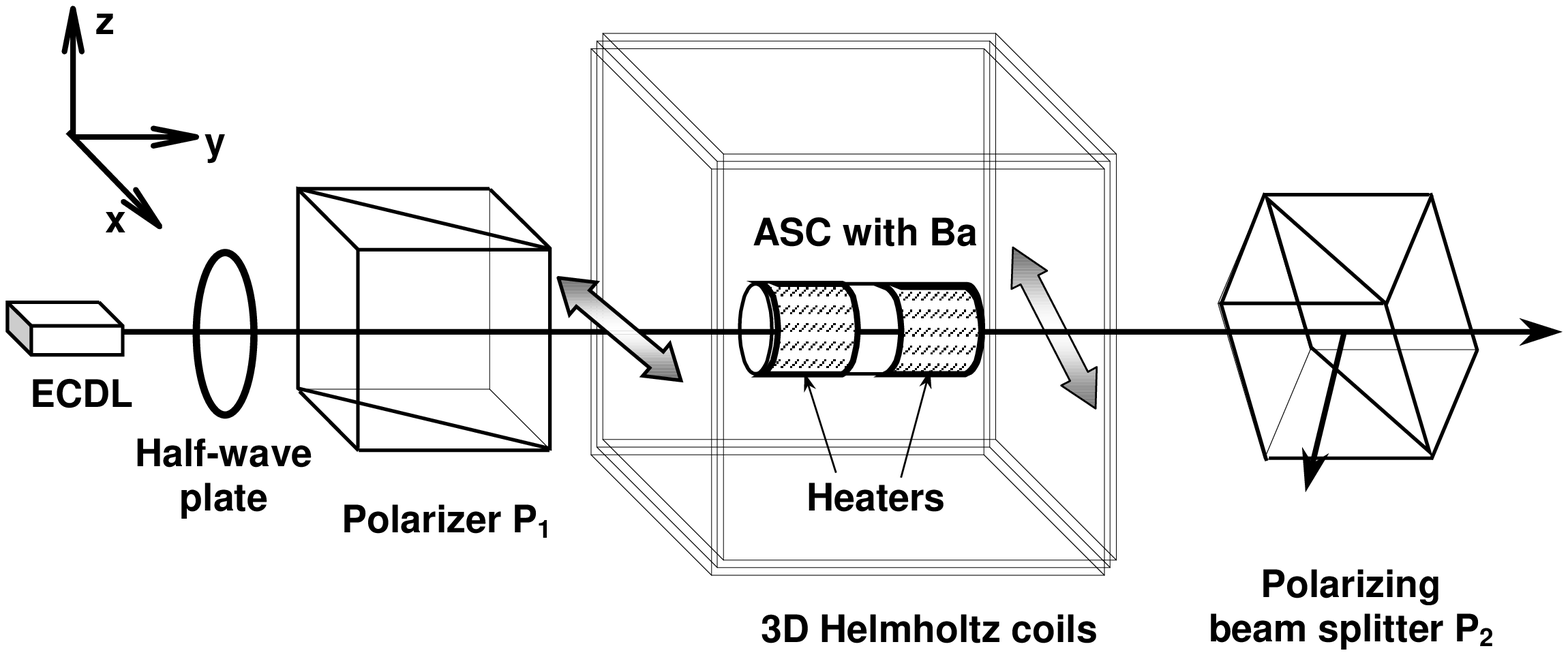}
\caption{\label{setup.fig}
    Scheme of the experimental apparatus.
}
\end{figure}

\subsection{Experimental configuration}

    A simplified scheme of the experimental setup is shown
in Fig.~\ref{setup.fig}.  We use an external cavity diode laser
(ECDL), tuned to the vicinity of the $6s^2~ {}^1S_0 \rightarrow
6s6p~ {}^1P_3$ intercombination transition of atomic Ba. The laser
beam passes through a high-quality polarizer $P_1$ (to insure the
quality of the linear polarization).  A half-wave plate placed in
front of the polarizer allows smooth control of the intensity of
the laser beam at the SC.  The cell (with the heaters and thermal
isolation) is mounted in the center of triaxial Helmholtz coils to
control the magnetic field within the interaction region.  The
system consists of three mutually orthogonal pairs of square coils
in Helmholtz configuration, with the magnetic field inhomogeneity
of $\pm 5$~mG within a $10$~cm cubical central region.

    After traversing the Ba cell, the transmission and the
polarization rotation are analyzed by using a polarization beam
splitter ($P_2$ in Fig.~\ref{setup.fig}) tilted at 45 degrees with
respect to the polarizer $P_1$.  In this configuration, the output
signal of the two channels of the beam splitter are given by
$S_{1,2}=\frac12 I_{\mathrm{out}}(1 \pm \sin 2\phi)$, where
$I_{\mathrm{out}}$ is the transmitted intensity of the laser beam,
and $\phi$ is the polarization rotation angle.

\section{Experimental results}
\subsection{Saturation absorption spectroscopy of Ba}

\begin{figure}
\centering
\includegraphics[width=1.00\columnwidth]{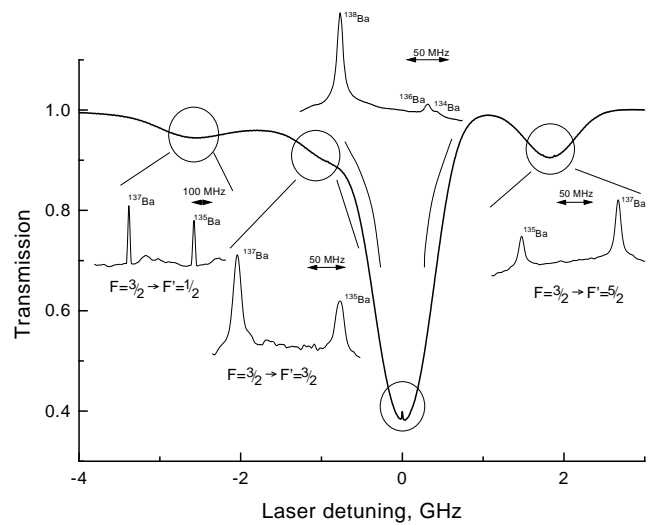}
\caption{\label{sat-abs.fig}
    Transmission spectrum of the laser field as its
    frequency is swept across the intercombination
    transition of Ba. \textit{Insets}: the saturation
    absorption spectra for each absorption peak. The power
    of the probe and pump fields are $150\mathrm{\mu W}$
    and $2\mathrm{mW}$ respectively. The atomic density
    of Ba is $N=1.8\times 10^{12}~\mathrm{cm}^{-3}$.
}
\end{figure}

    The transmission spectra of the laser is shown in
Fig.~\ref{sat-abs.fig}.  The central absorption peak is due to the
even isotopes of Ba, mainly ${}^{138}$Ba which accounts for $70\%$
of natural Ba with an addition of ${}^{136}$Ba ($6\%$),
${}^{134}$Ba ($2\%$) and ${}^{132}$Ba ($1\%$). Since the nuclear
spin for all even isotopes is zero, there is no hyperfine
structure.  Natural Ba also contains relatively small amount odd
isotopes ${}^{137}$Ba ($10\%$) and ${}^{135}$Ba ($10\%$), which
are responsible for the adjacent absorption peaks in
Fig.~\ref{sat-abs.fig}.  Since the nuclear spin of the odd
isotopes is $I=3/2$, we observe three resolved absorption lines
corresponding to the resonant transition to the excited hyperfine
states with $F'=1/2$, $3/2$, and $5/2$.

    It has been demonstrated that extremely narrow
resonances may be observed in Ba atoms using saturation
absorption spectroscopy~\cite{akulshin'92, ducloy'97}.
In our experiments, we also record Doppler-free spectra in
the Ba vapor cell, but mainly for isotope identification and
frequency calibration.

    The part of the laser beam reflected from the
polarizer $P_1$ is used as a counterpropagating pumping
beam.  The ratio between pump and probe field intensities
is controlled by the half-wave plate placed before $P_1$,
and the polarization of the pump beam may be changed by half-
or quater-wave plates placed after the polarizer.  The width
(FWHM) of the observed resonances is $\approx 8$~MHz, which
is determined by the spectral width of the laser radiation
and residual Doppler effect.  It is also interesting that the
amplitude of the observed resonances shows no dependence on
the polarization of the pumping beam.  The positions of the
transmission peaks are in very good agreement with previously
published data~\cite{grundevik'82}.

\begin{figure}
\centering
\includegraphics[width=0.70\columnwidth]{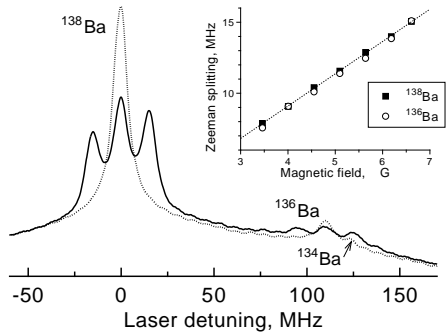}
\caption{\label{zeeman.fig}
    Saturated absorption resonances at zero magnetic
    field (dotted line) and $B=7.5~\mathrm{G}$ (solid
    line). \textit{Inset}:  Measured shift between the
    resonances as a function of magnetic field.
}
\end{figure}

    We also study the Zeeman shift of the magnetic sublevels
of the excited states in the presence of a longitudinal
magnetic field.  It is easy to calculate that for even
isotopes the gyromagnetic ratio of the ${}^3P_1$ state is
$\mathrm{g}=3/2$, which corresponds to a $2.11~\mathrm{MHz/G}$
shift of the magnetic sublevels~\cite{sobel'man_book}.  We have
measured the splitting of the saturation absorption resonances
as a function of applied magnetic field.  The results are
shown in Fig.~\ref{zeeman.fig}.  This method provides an
accurate calibration of the magnetic field in the system.
We also detected the Zeeman splitting for the transitions of
the odd isotopes.  However, individual magnetic resonances are
not clearly resolved there due to the rich magnetic substructure
of the hyperfine levels.

\subsection{Nonlinear magneto-optical effects}

\begin{figure}
\centering
\includegraphics[width=0.70\columnwidth]{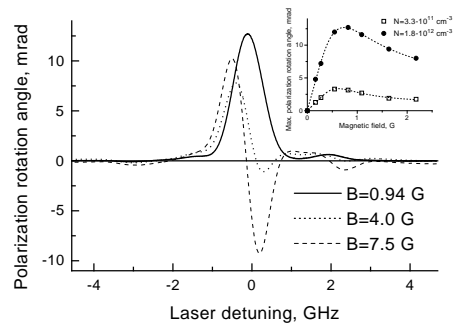}
\caption{\label{farad-rot.fig}
    The polarization rotation angle of linearly polarized
    light tuned across the Ba resonance for different
    values of the longitudinal magnetic field.  The laser
    power is $2.2$~mW, the density of Ba vapor is $1.8
    \times 10^{12}~\mathrm{cm}^{-3}$.  \textit{Inset}:
    maximum polarization rotation angle as a function of
    magnetic field.  Only the range of the magnetic fields
    where the nonlinear Faraday effect dominates is shown.
    The dotted lines are to guide the eyes.
}
\end{figure}

    Let us first consider the case when the magnetic
field is applied along the propagation direction of the
laser beam (Faraday configuration).  The spectra of the
polarization rotation angle for different magnetic fields
are shown in Fig.~\ref{farad-rot.fig}.  One can clearly see
two different regimes:  for smaller values of magnetic field
($B\leq 2~\mathrm{G}$) the rotation spectrum consists of one
peak, whereas for higher magnetic fields it becomes almost
antisymmetric with respect to the center of the absorption line.

\begin{figure}
\centering
\includegraphics[width=1.00\columnwidth]{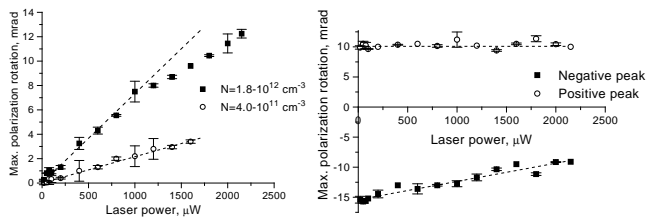}
\caption{\label{farad-rot-int.fig}
    (a) Maximum polarization rotation angle due to the
    nonlinear Faraday effect as a function of laser
    power. The data are taken at a magnetic field of
    $B=0.9$~G for two values of atomic density.  (b) The
    maximum values of the rotation peaks as functions
    of laser power for the linear interaction regime
    ($B=7.5$~G). The density of Ba vapor is $1.8\times
    10^{12}~\mathrm{cm}^{-3}$.
}
\end{figure}

    This behavior may be partially explained if we assume
that for small magnetic field the nonlinear magneto-optical
rotation is observed, and for higher $B$ the linear interaction
becomes dominant.  We can check this hypothesis by looking at the
dependance of the polarization rotation angle on the laser power
in both cases.  Nonlinear polarization rotation is caused by the
light-induced coherence between the $m=\pm 1$ sublevels of the
excited state, and therefore the magnitude of the rotation angle
should depend on the light intensity.  This will be shown in the
next Section. Measurements of the rotation angle, shown in
Fig.~\ref{farad-rot-int.fig}a are in good agreement with this
statement. This also means that in some cases (for example, when
the atomic transition is weak) it may be more convenient to study
atomic transitions with NMOE spectroscopy than by absorption
spectroscopy.

    The linear Faraday effect, in which the polarization
rotation is determined by the dispersion of the atomic transition,
should not demonstrate any dependence on laser intensity.
Fig.~\ref{farad-rot-int.fig}b shows the amplitudes of both
rotation peaks in the linear regime. One can see that the
amplitude of the positive peak does not change at all, whereas
that of negative peak reduces a little as the laser power
increases.  This variation may be due to residual nonlinear
Faraday effect. Overall, the behavior of the rotation angle
indicates that the polarization rotation for high magnetic field
is due to the linear interaction. However, there is no clear
explanation of the antisymmetric shape of the rotation spectra. It
is conceivable that it may be due to the influence of other even
isotopes, even though there is no clear physical reasons why they
cause the polarization rotation in the opposite direction, or of
the magnitude comparable with that of ${}^{138}$Ba.

\begin{figure}
\centering
\includegraphics[width=0.70\columnwidth]{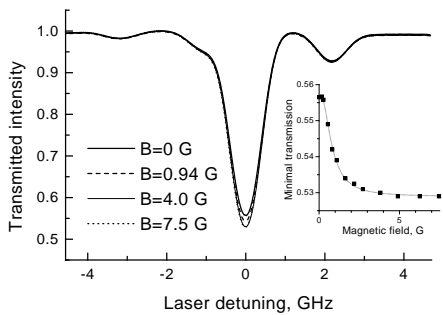}
\caption{\label{farad-abs.fig}
    Transmission through the cell for different values of
    magnetic field.  \textit{Inset:}  Laser transmission on
    resonance as a function of magnetic field.  Dotted line
    is a Lorenzian fit with FWHM of $1.7$~G. The laser
    power is $2.2$mW, the atomic density is $1.8 \times
    10^{12}~\mathrm{cm}^{-3}$.
}
\end{figure}

    It is also important to note that in the regime of
nonlinear interaction we observe suppression of absorption
for small values of magnetic field.  This is a manifestation
of Electromagnetically Induced Transparency~\cite{scullybook,
harris'97pt, marangos'98} due to the Zeeman coherence created
on the excited states magnetic sublevels.  The absorption
spectra for different values of magnetic field are shown in
Fig.~\ref{farad-abs.fig}.  One can also see that no change in
the laser absorption occurs in the linear interaction regime.

\begin{figure}
\centering
\includegraphics[width=0.70\columnwidth]{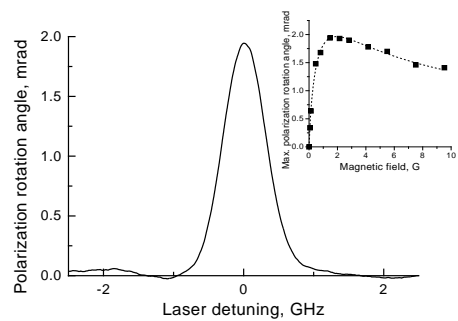}
\caption{\label{voigt-rot.fig}
    The polarization rotation angle of linearly polarized
    light as a function of the laser detuning for
    transverse magnetic field $B_z=2.0$~G. The laser
    power is $2.2$~mW, the density of Ba vapor is $1.8
    \times 10^{12}~\mathrm{cm}^{-3}$.  \textit{Inset}:
    maximum polarization rotation angle as a function of
    magnetic field.  The dotted lines are to guide the eyes.
}
\end{figure}

    We have also investigated the nonlinear magneto-optical
polarization rotation in the presence of a transverse magnetic
field.  The magnetic field is applied perpendicular to both
the electric component of the electromagnetic field and its
propagation direction (along the $z$ axis).  The polarization
rotation spectra is shown in Fig.~\ref{voigt-rot.fig}.
The magnitude of the polarization rotation is almost an order of
magnitude smaller than that for the nonlinear Faraday effect.
The other important difference is that the dependence of
the rotation angle on the magnetic field is symmetric in the
case of the transverse magnetic field, and asymmetric for the
longitudinal magnetic field (not shown in the graphs).

\begin{figure}
\centering
\includegraphics[width=0.60\columnwidth]{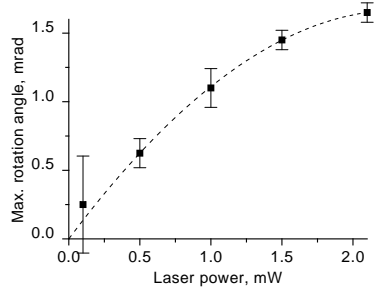}
\caption{\label{voigt-rot-int.fig}
    Maximum polarization rotation angle for the
    case of transverse magnetic field as a function of
    laser power.  The data are taken for magnetic field
    $B_z=1.8$~G.  The density of Ba vapor is $1.8 \times
    10^{12}~\mathrm{cm}^{-3}$.  The dotted line is to
    guide the eyes.
}
\end{figure}

    To verify the nonlinear nature of the observed
polarization rotation, we have measured the dependence of
the polarization rotation on the laser intensity.  As one
can see in Fig.~\ref{voigt-rot-int.fig}, the rotation angle
is proportional to the power of the electromagnetic field,
although this dependence is not linear.

    Because of the symmetry of the system, a magnetic
field applied parallel to the polarization direction of the
linearly polarized magnetic field (the $x$-axis) should not
produce any rotation.  In the experiment, we observe some
small ($<0.5$~mrad) rotation angle in the presence of magnetic
field along the $x$-axis, probably due to misalignment in the
experimental apparatus.

\section{Theoretical analysis}

\begin{figure}
\centering
\includegraphics[width=0.60\columnwidth]{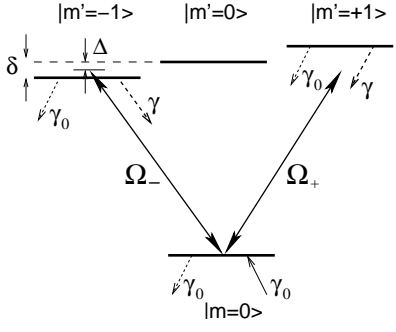}
\caption{\label{levels.fig}
    The interaction scheme of a linearly polarized
    electromagnetic field with Ba atoms in the presence
    of a longitudinal magnetic field.
}
\end{figure}

    The level scheme of the Ba transition is shown
in Fig.~\ref{levels.fig}.  Two components of the linearly
polarized light link the ground state $|m=0\rangle$ and two
excited states $|m'=\pm1\rangle$ into a $V$ configuration;
the third magnetic sublevel of the excited state cannot be
excited because of the selection rules.  We call the detuning
of the laser from the atomic resonance $\Delta$, and the
Zeeman splitting of the magnetic sublevels of the excited
states $\delta$.  Then the interaction Hamiltonian for this
system in the rotating wave approximation is given by:
\begin{eqnarray}    \label{ham}
H&=& \hbar (\Delta-\delta)|-\rangle\langle -| + \hbar
(\Delta+\delta)|+\rangle\langle +| \nonumber \\
&-& \hbar \left
(\Omega_+|+\rangle\langle 0| + \Omega_-|-\rangle\langle 0| + \mathrm{h.c.}\right)
\end{eqnarray}
Here and for the remainder of this discussion we let the
states $|\pm\rangle$ and $|0\rangle$ refer correspondingly
to the magnetic sublevels of the excited states $m'=\pm1$
and the ground state with $m=0$.

    Using the hamiltonian (\ref{ham}), we write the Bloch
equations for the density matrix elements
\begin{eqnarray}
\dot \rho_{00}&=&
\gamma_0(1-\rho_{00})+\gamma_r(\rho_{++}+\rho_{--}) + \nonumber
\\ &&i(\Omega_+^*\rho_{+0}+ \Omega_-^*\rho_{-0} - c.c.)
\label{rho00} \\
\dot\rho_{\pm\pm} &=& -(\gamma_0+\gamma_r+
\tilde\gamma_r)\rho_{\pm\pm}
-i(\Omega_\pm^*\rho_{\pm0}-\Omega_\pm\rho_{0\pm}) \label{rho++}\\
\dot\rho_{\pm0} &=&
-\Gamma_{\pm0}\rho_{\pm0}+i\Omega_\pm(\rho_{00}-\rho_{\pm\pm}) -
i\Omega_\mp\rho_{\pm\mp} \label{rho+0}\\
\dot\rho_{+-} &=& -\Gamma_{+-}\rho_{+-}+i\Omega_+\rho_{0-} -
i\Omega_+^*\rho_{+0}\label{rho+-}
\end{eqnarray}
where the polarization decay rates are given by:
\begin{eqnarray}
\Gamma_{\pm0}&=&(\gamma_0+\gamma_r/2+ \tilde\gamma_r/2)+i(\Delta \pm
\delta) \\
\Gamma_{+-}&=&(\gamma_0+\gamma_r+ \tilde\gamma_r)+i \cdot 2\delta
\end{eqnarray}
Here $\gamma_r$ is the radiative decay rate of the excited
states $m=\pm 1$ to the ground state, $\tilde\gamma_r$ is
the decay outside of the three-level system, $\gamma_0$
characterizes the finite interaction time of the atoms
with the laser beam.  In the following calculations, we
assume that $\gamma_0$ is small compared to $\gamma_r$,
and define $\gamma = \gamma_r + \tilde\gamma_r$ as the total
radiative decay rate of the excited states.  For simplicity,
we neglect the effect of spontaneous emission-induced coherent
effects~\cite{paspalakis'98}.

    After solving Eqs.~(\ref{rho00}-\ref{rho+-}) in the
steady-state regime, we obtain the following expression for
the polarization
\begin{equation} \label{rho0+gen}
\rho_{\pm0}=\frac{i\Omega_\pm}{\Gamma_{\pm0}}
\frac{(\Gamma_{\pm\mp}+\displaystyle{\frac{|\Omega_\pm|^2}{\Gamma_{0\mp}}})(\rho
_{00}-\rho_{++})
-\displaystyle{\frac{|\Omega_\mp|^2}{\Gamma_{0\mp}}}(\rho_{00}-\rho_{++})}
{\Gamma_{\pm\mp}+\displaystyle{\frac{|\Omega_\pm|^2}{\Gamma_{0\mp}}}+
\displaystyle{\frac{|\Omega_\mp|^2}{\Gamma_{\pm0}}}} ~.
\end{equation}
Let us first briefly consider the linear interaction regime, such
that magneto-optical effects do not depend on the intensity of the
electromagnetic field and are determined solely by the resonant
absorption and dispersion of the two-level interaction.  From
Eq.~(\ref{rho0+gen}) it is easy to see that this regime is
observed either for low intensity of the laser field
($|\Omega_\pm| \ll \gamma$), or for strong magnetic field ($\delta
\gg |\Omega_\pm|,|\Omega_\pm|^2/\gamma$). In this case no
coherence between excited states is created. The populations of
the excited magnetic sublevels are determined by Rabi frequencies
of the corresponding circularly polarized electro-magnetic fields,
and are small compared to the ground-state population.  Thus, the
propagation equation for the circular components of the laser
field is:
\begin{equation} \label{eprop-lin}
\frac{\partial\Omega_\pm}{\partial z} = - \frac{\kappa
\Omega_\pm}{\gamma/2 +i(\Delta \pm \delta)}
\end{equation}
where $\kappa = \frac{3}{8\pi}N\lambda^2\gamma_r$, and
$|\Omega|^2 = |\Omega_+|^2+|\Omega_-|^2$ is the total
intensity of the electromagnetic field.  Substituting
$\Omega_\pm = |\Omega_\pm|\exp{i\phi_pm}$ and taking the real
part of Eq.~(\ref{eprop-lin}) we arrive at expressions
for the absorption coefficient $\alpha$ (defined as
$|\Omega(L)|^2=|\Omega(0)|^2e^{-\alpha L}$, where $|\Omega(L)|$
and $|\Omega(0)|$ are the values of the Rabi frequencies of
the laser field before and after the cell)
\begin{equation} \label{abs_lin}
\alpha = \kappa \gamma
\frac{\gamma^2/4+\Delta^2+\delta^2}{(\gamma^2/4+\Delta^2+\delta^2)^2-4\Delta^2\delta^2}
\end{equation}
The polarization rotation angle is given by half the
difference between the acquired phases of the circularly
polarized fields $\phi = (\phi_+-\phi_-)/2$. In case of the
linear Faraday effect it is equal to
\begin{equation} \label{rot_lin}
\phi = \kappa \delta \L
\frac{\gamma^2/4-\Delta^2+\delta^2}{(\gamma^2/4+\Delta^2+\delta^2)^2-4\Delta^2\delta^2}
\end{equation}

    As one can see, Eq.~(\ref{rot_lin}) is symmetric
with respect to the laser detuning $\Delta$ and therefore
does not explain the dispersion-like curves observed in
the experiment.  However, this theoretical analysis is not
complete without taking into account the velocity distribution
of moving Ba atoms.  It is known that in the hot vapor the
atomic velocities are described by a Maxwell distribution:
$dN\{v\}=N\exp{(-v^2/v_T^2)}dv$, where $v_T=\sqrt{2k_BT/m}$
is the most probable velocity, $k_b$ is the Boltzmann constant,
$T$ is the temperature of atomic vapor in K, and $m$ is the mass
of a Ba atom.  Because of the Doppler effect, the atoms from
different velocity groups resonate at different frequencies.
To account for this in our calculations it is necessary to
modify Eq.~(\ref{eprop-lin}) by replacing the one-photon
detuning $\Delta$ with $\Delta + kv$, where $k=2\pi/\lambda$
is the wave vector of the electromagnetic field, and then
averaging the right-hand side of Eq.~(\ref{eprop-lin}) over
the velocity distribution.

    The nonlinear magneto-optical effects appear due to the
coherence which may be created among the Zeeman sublevels of the
excited state.  Unlike a three-level $\Lambda$ system, where
the ground-state coherence arises from optical pumping of the
atoms in the noninteracting superposition of two ground-states
(the ``dark state''), in a $V$ scheme the coherence between the
two excited states is formed by the interference of different
excitation passes~\cite{scullybook}.

    In the approximation of strong electromagnetic field
($|\Omega| \gg \gamma$) we may assume that the population
distribution in the system is determined mostly by the field.
Therefore, we can assume that the atom is in one of the
eigenstates of the Hamiltonian (\ref{ham}), which allow
us to find the values of the populations of all levels in
zeroth approximation.  It is quite easy to find this state
for $\delta=0$:
\begin{equation}    \label{eigenfunc}
|\Psi\rangle = \mathcal{N}\left \{\Omega _+^*|+\rangle + \Omega _-^*|-\rangle -
\left(\frac{\Delta}{2}+\sqrt{\frac{\Delta^2}{4}+|\Omega|^2}\right)|0\rangle\right \}
\end{equation}
where the normalization coefficient $\mathcal{N}$ is given by
\begin{equation}  \label{norm}
\frac{1}{\mathcal{N}^2}=  2
\sqrt{\frac{\Delta^2}{4}+|\Omega|^2}\left(\frac{\Delta}{2}+\sqrt{\frac{\Delta^2}
{4}+|\Omega|^2}\right). \end{equation} In this case we find the population
differences between the excited states and the ground state are the following:
\begin{equation} \label{pop-zero}
\rho_{00}^{(0)}-\rho_{\pm\pm}^{(0)} = \mathcal{N}^2\left\{\frac{\Delta^2}{2}+\Delta\sqrt{\frac{\Delta^2}{4}+|\Omega|^2}+|\Omega_\mp|^2\right\} ~.
\end{equation}
Substituting the expressions for atomic populations above
into Eq.~(\ref{rho0+gen}) we may find the values of atomic
polarizations in the approximation $|\Omega| \gg \delta,
\gamma, \Delta$
\begin{equation} \label{rho_0+}
\rho_{\pm0} \simeq i\Omega_\pm\frac{(\gamma+2i\delta-2i\Delta)|\Omega_\mp|^2}{|\Omega|^4} ~.
\end{equation}
Using Eq.~(\ref{rho_0+}) it is easy to calculate the transmitted
intensity and the polarization angle of the linearly polarized
electromagnetic field
\begin{eqnarray}    \label{abs_nlin}
|\Omega(L)|^2 &=& |\Omega(0)|^2 - \kappa\gamma\L \\
\phi &=& \frac{2\delta}{\gamma}\ln\frac{|\Omega(0)|^2}{|\Omega(L)|^2}
\label{rot_nlin}
\end{eqnarray}
This result is in a good agreement with the previous calculations
~\cite{schuller'87} in the limit of optically thin medium. Note
that this expression is very similar to that for the $\Lambda$
system~\cite{matsko'03} with the decay rate of the ground-state
coherence replaced by the radiative decay of the excited states.

    From Eq.~(\ref{rot_nlin}) one can see that the
polarization rotation angle is inversely proportional to the
radiative width of the excited state.  Therefore, it is clear
that several orders of magnitude may be gained in polarization
rotation and other nonlinear effects in $V$ interaction scheme
by using narrow alkaline-earth intercombination transitions
compared with traditional alkali atoms.

    It is easy to see, however, that the theory developed
above predicts much larger values of the polarization rotation
angle than experimentally observed.  There are several effects
which may contribute to this discrepancy.  For example, we do not
to take into account thermal motion of atoms which, from one hand,
leads to the inhomogeneous Doppler broadening, and from the other
hand - limits the interaction time of Ba atoms with light. Indeed,
the average thermal speed of Ba atoms at about $500^o$~C is about
$300$~m/s, which corresponds to a decay rate $\gamma_0 \approx
2\pi \times 50$~kHz and is comparable with the radiative width of
the transition. This may change the population distribution
between atomic levels. Another important factor is the presence of
relatively dense ($>10^{13}~\mathrm{cm}^{-3}$) Rb vapor due to
small contamination of Ba sample, since it has been demonstrated
that collisions cause the reduction of the polarization
rotation~\cite{schuller'87}.

\section{Summary}

    In this paper, we have presented an experimental study
of linear and nonlinear optical effects on the intercombination
transition 6S-6P of Ba atoms in sealed vapor cell.  We study
the Zeeman shift of the magnetic sublevels of the excited state
and measure the polarization rotation spectra in presence of
longitudinal and transverse magnetic fields.  These nonlinear
optical effects can have a wide range of applications as a
spectroscopic tool with good resolution and sensitivity for
studying weak atomic transitions.

\section{Acknowledgements}

    The authors would like to thank E.\ E.\ Mikhailov,
Y.\ Malakyan, V.\ A.\ Sautenkov, and A.\ S.\ Zibrov for useful
and stimulating discussions, and Office of Naval Research for
financial support.


\bibliography{ba-nmor}

\begin{thebibliography}{26}
\expandafter\ifx\csname natexlab\endcsname\relax\def\natexlab#1{#1}\fi
\expandafter\ifx\csname bibnamefont\endcsname\relax
  \def\bibnamefont#1{#1}\fi
\expandafter\ifx\csname bibfnamefont\endcsname\relax
  \def\bibfnamefont#1{#1}\fi
\expandafter\ifx\csname citenamefont\endcsname\relax
  \def\citenamefont#1{#1}\fi
\expandafter\ifx\csname url\endcsname\relax
  \def\url#1{\texttt{#1}}\fi
\expandafter\ifx\csname urlprefix\endcsname\relax\def\urlprefix{URL }\fi
\providecommand{\bibinfo}[2]{#2}
\providecommand{\eprint}[2][]{\url{#2}}

\bibitem[{\citenamefont{Budker et~al.}(2002)\citenamefont{Budker, Gawlik,
  Kimball, Rochester, Yashchuk, and Weis}}]{budker'02rmp}
\bibinfo{author}{\bibfnamefont{D.}~\bibnamefont{Budker}},
  \bibinfo{author}{\bibfnamefont{W.}~\bibnamefont{Gawlik}},
  \bibinfo{author}{\bibfnamefont{D.~F.} \bibnamefont{Kimball}},
  \bibinfo{author}{\bibfnamefont{S.~M.} \bibnamefont{Rochester}},
  \bibinfo{author}{\bibfnamefont{V.~V.} \bibnamefont{Yashchuk}},
  \bibnamefont{and} \bibinfo{author}{\bibfnamefont{A.}~\bibnamefont{Weis}},
  \bibinfo{journal}{Rev. Mod. Phys.} \textbf{\bibinfo{volume}{74}},
  \bibinfo{pages}{1153} (\bibinfo{year}{2002}).

\bibitem[{\citenamefont{Drake et~al.}(1988)\citenamefont{Drake, Lange, and
  Mlynek}}]{drake'88}
\bibinfo{author}{\bibfnamefont{K.~H.} \bibnamefont{Drake}},
  \bibinfo{author}{\bibfnamefont{W.}~\bibnamefont{Lange}}, \bibnamefont{and}
  \bibinfo{author}{\bibfnamefont{J.}~\bibnamefont{Mlynek}},
  \bibinfo{journal}{Opt. Commun.} \textbf{\bibinfo{volume}{66}},
  \bibinfo{pages}{315} (\bibinfo{year}{1988}).

\bibitem[{\citenamefont{Barkov et~al.}(1989)\citenamefont{Barkov,
  Melik-Pashaev, and Zolotorev}}]{barkov'89}
\bibinfo{author}{\bibfnamefont{L.~M.} \bibnamefont{Barkov}},
  \bibinfo{author}{\bibfnamefont{D.~A.} \bibnamefont{Melik-Pashaev}},
  \bibnamefont{and} \bibinfo{author}{\bibfnamefont{M.~S.}
  \bibnamefont{Zolotorev}}, \bibinfo{journal}{Opt. Commun.}
  \textbf{\bibinfo{volume}{70}}, \bibinfo{pages}{467} (\bibinfo{year}{1989}).

\bibitem[{\citenamefont{Agarwal et~al.}(1997)\citenamefont{Agarwal, Lakshmi,
  Connerade, and West}}]{agarwal'97}
\bibinfo{author}{\bibfnamefont{G.~S.} \bibnamefont{Agarwal}},
  \bibinfo{author}{\bibfnamefont{P.~A.} \bibnamefont{Lakshmi}},
  \bibinfo{author}{\bibfnamefont{J.~P.} \bibnamefont{Connerade}},
  \bibnamefont{and} \bibinfo{author}{\bibfnamefont{S.}~\bibnamefont{West}},
  \bibinfo{journal}{J. Phys. B} \textbf{\bibinfo{volume}{30}},
  \bibinfo{pages}{5971} (\bibinfo{year}{1997}).

\bibitem[{\citenamefont{Schuller et~al.}(1987)\citenamefont{Schuller,
  Macpherson, and Stacey}}]{schuller'87}
\bibinfo{author}{\bibfnamefont{F.}~\bibnamefont{Schuller}},
  \bibinfo{author}{\bibfnamefont{M.~J.~D.} \bibnamefont{Macpherson}},
  \bibnamefont{and} \bibinfo{author}{\bibfnamefont{D.~N.}
  \bibnamefont{Stacey}}, \bibinfo{journal}{Physica C}
  \textbf{\bibinfo{volume}{147}}, \bibinfo{pages}{321} (\bibinfo{year}{1987}).

\bibitem[{\citenamefont{Schuller et~al.}(1992)\citenamefont{Schuller,
  Warrington, Zetie, Macpherson, and Stacey}}]{schuller'92}
\bibinfo{author}{\bibfnamefont{F.}~\bibnamefont{Schuller}},
  \bibinfo{author}{\bibfnamefont{R.~B.} \bibnamefont{Warrington}},
  \bibinfo{author}{\bibfnamefont{K.~P.} \bibnamefont{Zetie}},
  \bibinfo{author}{\bibfnamefont{M.~J.~D.} \bibnamefont{Macpherson}},
  \bibnamefont{and} \bibinfo{author}{\bibfnamefont{D.~N.}
  \bibnamefont{Stacey}}, \bibinfo{journal}{Opt. Commun.}
  \textbf{\bibinfo{volume}{93}}, \bibinfo{pages}{169} (\bibinfo{year}{1992}).

\bibitem[{\citenamefont{Schuller and Stacey}(1999)}]{schuller'99}
\bibinfo{author}{\bibfnamefont{F.}~\bibnamefont{Schuller}} \bibnamefont{and}
  \bibinfo{author}{\bibfnamefont{D.~N.} \bibnamefont{Stacey}},
  \bibinfo{journal}{Phys. Rev. A} \textbf{\bibinfo{volume}{60}},
  \bibinfo{pages}{973} (\bibinfo{year}{1999}).

\bibitem[{\citenamefont{Zibrov et~al.}(1995)\citenamefont{Zibrov, Lukin,
  Nikonov, Hollberg, Scully, Velichansky, and Robinson}}]{zibrov'95}
\bibinfo{author}{\bibfnamefont{A.~S.} \bibnamefont{Zibrov}},
  \bibinfo{author}{\bibfnamefont{M.~D.} \bibnamefont{Lukin}},
  \bibinfo{author}{\bibfnamefont{D.~E.} \bibnamefont{Nikonov}},
  \bibinfo{author}{\bibfnamefont{L.}~\bibnamefont{Hollberg}},
  \bibinfo{author}{\bibfnamefont{M.~O.} \bibnamefont{Scully}},
  \bibinfo{author}{\bibfnamefont{V.~L.} \bibnamefont{Velichansky}},
  \bibnamefont{and} \bibinfo{author}{\bibfnamefont{H.~G.}
  \bibnamefont{Robinson}}, \bibinfo{journal}{Phys. Rev. Lett.}
  \textbf{\bibinfo{volume}{75}}, \bibinfo{pages}{1499} (\bibinfo{year}{1995}).

\bibitem[{\citenamefont{Grynberg et~al.}(1996)\citenamefont{Grynberg, Pinard,
  and Mandel}}]{grynberg'96}
\bibinfo{author}{\bibfnamefont{G.}~\bibnamefont{Grynberg}},
  \bibinfo{author}{\bibfnamefont{M.}~\bibnamefont{Pinard}}, \bibnamefont{and}
  \bibinfo{author}{\bibfnamefont{P.}~\bibnamefont{Mandel}},
  \bibinfo{journal}{Phys.\ Rev. A} \textbf{\bibinfo{volume}{54}},
  \bibinfo{pages}{776} (\bibinfo{year}{1996}).

\bibitem[{\citenamefont{Zhou and Swain}(1997)}]{zhou'97}
\bibinfo{author}{\bibfnamefont{P.}~\bibnamefont{Zhou}} \bibnamefont{and}
  \bibinfo{author}{\bibfnamefont{S.}~\bibnamefont{Swain}},
  \bibinfo{journal}{Phys. Rev. Lett.} \textbf{\bibinfo{volume}{78}},
  \bibinfo{pages}{832} (\bibinfo{year}{1997}).

\bibitem[{\citenamefont{Akulshin et~al.}(1998)\citenamefont{Akulshin, Barreiro,
  and Lezama}}]{akulshin'98}
\bibinfo{author}{\bibfnamefont{A.~M.} \bibnamefont{Akulshin}},
  \bibinfo{author}{\bibfnamefont{S.}~\bibnamefont{Barreiro}}, \bibnamefont{and}
  \bibinfo{author}{\bibfnamefont{A.}~\bibnamefont{Lezama}},
  \bibinfo{journal}{Phys.\ Rev.\ A} \textbf{\bibinfo{volume}{57}},
  \bibinfo{pages}{2996 } (\bibinfo{year}{1998}).

\bibitem[{\citenamefont{Taichenachev et~al.}(2000)\citenamefont{Taichenachev,
  Tumaikin, and Yudin}}]{taichenachev'00pra}
\bibinfo{author}{\bibfnamefont{A.~V.} \bibnamefont{Taichenachev}},
  \bibinfo{author}{\bibfnamefont{A.~M.} \bibnamefont{Tumaikin}},
  \bibnamefont{and} \bibinfo{author}{\bibfnamefont{V.~I.} \bibnamefont{Yudin}},
  \bibinfo{journal}{Phys. Rev. A} \textbf{\bibinfo{volume}{61}},
  \bibinfo{pages}{011802} (\bibinfo{year}{2000}).

\bibitem[{\citenamefont{Alzetta et~al.}(2001)\citenamefont{Alzetta, Cartaleva,
  Dancheva, Andreeva, Gozzini, Botti, and Rossi}}]{alzetta'01}
\bibinfo{author}{\bibfnamefont{G.}~\bibnamefont{Alzetta}},
  \bibinfo{author}{\bibfnamefont{S.}~\bibnamefont{Cartaleva}},
  \bibinfo{author}{\bibfnamefont{Y.}~\bibnamefont{Dancheva}},
  \bibinfo{author}{\bibfnamefont{C.}~\bibnamefont{Andreeva}},
  \bibinfo{author}{\bibfnamefont{S.}~\bibnamefont{Gozzini}},
  \bibinfo{author}{\bibfnamefont{L.}~\bibnamefont{Botti}}, \bibnamefont{and}
  \bibinfo{author}{\bibfnamefont{A.}~\bibnamefont{Rossi}}, \bibinfo{journal}{J.
  Opt. B} \textbf{\bibinfo{volume}{3}}, \bibinfo{pages}{181 }
  (\bibinfo{year}{2001}).

\bibitem[{\citenamefont{Nesmeyanov}(1963)}]{nesmeyanov'63}
\bibinfo{author}{\bibfnamefont{A.~N.} \bibnamefont{Nesmeyanov}},
  \emph{\bibinfo{title}{Vapour Pressure of the Elements}}
  (\bibinfo{publisher}{Academic Press, New York}, \bibinfo{year}{1963}).

\bibitem[{\citenamefont{Hinnov and Ohlendorf}(1969)}]{hinnov'69}
\bibinfo{author}{\bibfnamefont{E.}~\bibnamefont{Hinnov}} \bibnamefont{and}
  \bibinfo{author}{\bibfnamefont{W.}~\bibnamefont{Ohlendorf}},
  \bibinfo{journal}{J. Chem. Phys.} \textbf{\bibinfo{volume}{50}},
  \bibinfo{pages}{3005} (\bibinfo{year}{1969}).

\bibitem[{\citenamefont{Alcock et~al.}(1984)\citenamefont{Alcock, Itkin, and
  Horrigan}}]{alcock'84}
\bibinfo{author}{\bibfnamefont{C.~B.} \bibnamefont{Alcock}},
  \bibinfo{author}{\bibfnamefont{V.~P.} \bibnamefont{Itkin}}, \bibnamefont{and}
  \bibinfo{author}{\bibfnamefont{M.~K.} \bibnamefont{Horrigan}},
  \bibinfo{journal}{Canadian Metallurgical Quarterly}
  \textbf{\bibinfo{volume}{23}}, \bibinfo{pages}{309} (\bibinfo{year}{1984}).

\bibitem[{\citenamefont{Jacob and Waseda}(1988)}]{jacob'88}
\bibinfo{author}{\bibfnamefont{K.~T.} \bibnamefont{Jacob}} \bibnamefont{and}
  \bibinfo{author}{\bibfnamefont{Y.}~\bibnamefont{Waseda}},
  \bibinfo{journal}{J. of the Less-Common Metals}
  \textbf{\bibinfo{volume}{139}}, \bibinfo{pages}{249} (\bibinfo{year}{1988}).

\bibitem[{\citenamefont{Akulshin et~al.}(1992)\citenamefont{Akulshin, Celikov,
  and Velichansky}}]{akulshin'92}
\bibinfo{author}{\bibfnamefont{A.~M.} \bibnamefont{Akulshin}},
  \bibinfo{author}{\bibfnamefont{A.~A.} \bibnamefont{Celikov}},
  \bibnamefont{and} \bibinfo{author}{\bibfnamefont{V.~L.}
  \bibnamefont{Velichansky}}, \bibinfo{journal}{Opt. Commun.}
  \textbf{\bibinfo{volume}{93}}, \bibinfo{pages}{54} (\bibinfo{year}{1992}).

\bibitem[{\citenamefont{Loe-Mie et~al.}(1997)\citenamefont{Loe-Mie, Papoyan,
  Akulshin, Lazema, Leite, Lopez, Bloch, and Ducloy}}]{ducloy'97}
\bibinfo{author}{\bibfnamefont{R.}~\bibnamefont{Loe-Mie}},
  \bibinfo{author}{\bibfnamefont{A.~V.} \bibnamefont{Papoyan}},
  \bibinfo{author}{\bibfnamefont{A.~M.} \bibnamefont{Akulshin}},
  \bibinfo{author}{\bibfnamefont{A.}~\bibnamefont{Lazema}},
  \bibinfo{author}{\bibfnamefont{J.~R.~R.} \bibnamefont{Leite}},
  \bibinfo{author}{\bibfnamefont{O.}~\bibnamefont{Lopez}},
  \bibinfo{author}{\bibfnamefont{D.}~\bibnamefont{Bloch}}, \bibnamefont{and}
  \bibinfo{author}{\bibfnamefont{M.}~\bibnamefont{Ducloy}},
  \bibinfo{journal}{Opt. Commun.} \textbf{\bibinfo{volume}{139}},
  \bibinfo{pages}{55} (\bibinfo{year}{1997}).

\bibitem[{\citenamefont{Grundevik et~al.}(1983)\citenamefont{Grundevik,
  Gustavsson, Olsson, and Olsson}}]{grundevik'82}
\bibinfo{author}{\bibfnamefont{P.}~\bibnamefont{Grundevik}},
  \bibinfo{author}{\bibfnamefont{M.}~\bibnamefont{Gustavsson}},
  \bibinfo{author}{\bibfnamefont{G.}~\bibnamefont{Olsson}}, \bibnamefont{and}
  \bibinfo{author}{\bibfnamefont{T.}~\bibnamefont{Olsson}},
  \bibinfo{journal}{Z. Phys. A} \textbf{\bibinfo{volume}{312}},
  \bibinfo{pages}{1} (\bibinfo{year}{1983}).

\bibitem[{\citenamefont{Sobel'man}(1972)}]{sobel'man_book}
\bibinfo{author}{\bibfnamefont{I.~I.} \bibnamefont{Sobel'man}},
  \emph{\bibinfo{title}{Introduction to the theory of atomic spectra}}
  (\bibinfo{publisher}{Pergamon Press, Oxford}, \bibinfo{year}{1972}).

\bibitem[{\citenamefont{Scully and Zubairy}(1997)}]{scullybook}
\bibinfo{author}{\bibfnamefont{M.~O.} \bibnamefont{Scully}} \bibnamefont{and}
  \bibinfo{author}{\bibfnamefont{M.~S.} \bibnamefont{Zubairy}},
  \emph{\bibinfo{title}{Quantum Optics}} (\bibinfo{publisher}{Cambridge
  University Press, Cambridge, UK}, \bibinfo{year}{1997}).

\bibitem[{\citenamefont{Harris}(1997)}]{harris'97pt}
\bibinfo{author}{\bibfnamefont{S.~E.} \bibnamefont{Harris}},
  \bibinfo{journal}{Phys.\ Today} \textbf{\bibinfo{volume}{50}},
  \bibinfo{pages}{36} (\bibinfo{year}{1997}).

\bibitem[{\citenamefont{Marangos}(1998)}]{marangos'98}
\bibinfo{author}{\bibfnamefont{J.~P.} \bibnamefont{Marangos}},
  \bibinfo{journal}{J.\ Mod.\ Opt.} \textbf{\bibinfo{volume}{45}},
  \bibinfo{pages}{471 } (\bibinfo{year}{1998}).

\bibitem[{\citenamefont{Paspalakis et~al.}(1998)\citenamefont{Paspalakis, Gong,
  and Knight}}]{paspalakis'98}
\bibinfo{author}{\bibfnamefont{E.}~\bibnamefont{Paspalakis}},
  \bibinfo{author}{\bibfnamefont{S.-Q.} \bibnamefont{Gong}}, \bibnamefont{and}
  \bibinfo{author}{\bibfnamefont{P.~L.} \bibnamefont{Knight}},
  \bibinfo{journal}{Opt. Commun.} \textbf{\bibinfo{volume}{152}},
  \bibinfo{pages}{293} (\bibinfo{year}{1998}).

\bibitem[{\citenamefont{Matsko et~al.}(2003)\citenamefont{Matsko, Novikova,
  Zubairy, and Welch}}]{matsko'03}
\bibinfo{author}{\bibfnamefont{A.~B.} \bibnamefont{Matsko}},
  \bibinfo{author}{\bibfnamefont{I.}~\bibnamefont{Novikova}},
  \bibinfo{author}{\bibfnamefont{M.~S.} \bibnamefont{Zubairy}},
  \bibnamefont{and} \bibinfo{author}{\bibfnamefont{G.~R.} \bibnamefont{Welch}},
  \bibinfo{journal}{Phys. Rev. A} \textbf{\bibinfo{volume}{67}},
  \bibinfo{pages}{043805} (\bibinfo{year}{2003}).

\end{thebibliography}

\end{document}